\documentclass{article}
\title{Berry phase for coherent states in spin systems }
\author{Kh.Kh.Muminov, Y. Yousefi \\
 Physical-Technical Institute named after S.U.Umarov \\
 Academy of Sciences of the Republic of Tajikistan  \\
 Aini Ave 299/1, Dushanbe, Tajikistan, \\
 e-mail: khikmat@inbox.ru, muminov@phti.th}
\date{}
\begin{document}
\maketitle
\begin{abstract}
In this paper we obtain Berry phase from Schrödinger equation. For vector states, basic kets are coherent states in real parameterization. We calculate Berry phase for spin S=1/2 and spin S=1 in SU(2) group and  Berry phase for spin S=1 in SU(3) group.
\end{abstract}
\section{Introduction}
In mechanics (including classical mechanics as well as quantum mechanics), the Geometric phase, or the Pancharatnam-Berry phase (named after S. Pancharatnam and Sir Michael Berry), also known as the Pancharatnam phase or, more commonly, Berry phase[1,2], is a phase acquired over the course of a cycle, when the system is subjected to cyclic adiabatic processes, resulting from the geometrical properties of the parameter space of the Hamiltonian. Apart from quantum mechanics, it arises in a variety of other wave systems, such as classical optics [3]. As a rule of thumb, it occurs whenever there are at least two parameters affecting a wave, in the vicinity of some sort of singularity or some sort of hole in the topology. In nonrelativistic quantum mechanics, the state of a system is described by the vector of the Hilbert space (the wave function) $\psi\in H$ which depends on time and some set of other variables depending on the considered problem. The evolution of a quantum system in time t is described by the Schrodinger equation 

\begin{eqnarray}
i\hbar \frac{\partial \psi}{\partial t}=H\psi
\end{eqnarray}
where H is called Hamiltonian of a system, and $\hbar$  is the Planck constant. For simplicity, we put  and denote partial derivative on time by a dot, $\dot \psi =\partial_t \psi$, in what follows.  

Assume that, the state vectors to be normalized on unit, 

\begin{eqnarray}
\parallel \psi \parallel= \sqrt{\langle \psi | \psi\rangle}=1
\end{eqnarray}

The norm of a state vector is preserved in time due to the self-adjointness of the Hamiltonian. Normalization of a state vector does not eliminate the arbitrariness in choosing the state vector in the Hilbert space because the arbitrariness in choosing a phase factor still remains.

Now we describe the problem which was considered by M. Berry [4] in its simplest form. Assume for simplicity that the Hilbert space is a finite dimensional and a state vector is represented by a column of N components,

\begin{eqnarray}
\psi=
\left (
\begin{array}{c}
\psi_1 \\ \vdots \\ \psi_N
\end{array}
\right )
\end{eqnarray}
where $ \psi_1,...,\psi_N$  are complex valued functions on some set of variables which will be specified later . 

Any solution of the Schrodinger equation (1) with normalization condition (2) is defined up to a constant phase factor $ e^{i\Theta_0}, \Theta_0=cte$ .
We consider the eigenvalue problem

\begin{eqnarray}
H\phi=E\phi, & & {  }E=cte
\end{eqnarray}
where $\phi\in H$ is coherent state . Suppose there exists nondegenerate energy eigenvalue E which depends on $\lambda$  differentiably. The eigenfunction $ \phi(\lambda)$  is also assumed to be a differentiable function on $ \lambda$  . Without loss of generality, we suppose that the eigenfunction $ \phi$  is normalized on unit, $\langle \phi | \phi \rangle=1$  . Then it is unique up to a multiplication on a phase factor which may be $\lambda$   dependent.

In the adiabatic approximation, for slowly varying Hamiltonian, the system remains in its instantaneous eigenstate. Therefore we look for the solution in the form

\begin{eqnarray}
\psi=e^{i\Theta}\phi
\end{eqnarray}
where $\Theta=\Theta(\lambda)$  is an unknown function on $ \lambda$ . Substitution of this expression into the Schrodinger equation yields the equation for the phase factor

\begin{eqnarray}
-\dot \Theta \phi+i\dot \phi=E\phi
\end{eqnarray}
where we dropped the common phase factor $e^{i\Theta}$  and used the commutative of matrices $He^{i\Theta}=e^{i\Theta}H$  . Now we take the scalar product of left and right sides of the derived equation with $\phi$ . As a result, we obtain the equation for the phase

\begin{eqnarray}
\dot \Theta =i\langle \phi|\dot \phi \rangle -E
\end{eqnarray}
and initial condition is $ \phi|_{t=0}=0$ .  Since $\dot \phi=\dot \lambda^k \partial_k \phi$ , the solution for equation (7) is

\begin{eqnarray}
\Theta=\int_0^t dt\dot \lambda^k A_k -\int_0^t dtE= \int_{\lambda(0)}^{\lambda(t)} d\lambda^k A_k-\int_0^t dsE(s)
\end{eqnarray}
where we introduced the notation 

\begin{eqnarray}
A_k(\lambda)=i\langle \phi| \partial_k \phi \rangle
\end{eqnarray}
and the integral on $\lambda$   is taken along the curve $\lambda(t)$ .

The first term in Eq.(8) is called the geometric or Berry’s phase and the second term is called the dynamical phase.

Note that components (9) are real because of normalization of the wave function. Indeed, differentiation of the normalization condition  $\langle \phi | \phi \rangle=1$   yields the equality

\begin{eqnarray}
\langle \partial_k \phi | \phi \rangle+\langle  \phi |\partial_k \phi \rangle=\langle  \phi |\partial_k \phi \rangle^++\langle  \phi |\partial_k \phi \rangle=0
\end{eqnarray}

It implies the reality of components (9) and subsequently the reality of Berry phase.

Then the total change in the phase of the wave function is equal to the integral

\begin{eqnarray}
\Theta=\Theta_B-\int_0^t dtE
\end{eqnarray}
where

\begin{eqnarray}
\Theta_B=\oint_{\lambda}  d\lambda^k A_k
\end{eqnarray}

In this form, we are able to give the geometrical interpretation of the Berry phase $\Theta_B$   which is given by the first term in the obtained expression.

The expression for the Berry phase (12) can be rewritten as a surface integral of the components of the local curvature form. Using Stoke’s formulae, we obtain the following expression

\begin{eqnarray}
\Theta_B=\frac{1}{2}\int \int_s  d\lambda^k \times d\lambda^l F_{kl}
\end{eqnarray}
where S is a surface in $ R^3$  and $ F^{kl}=\partial_k A_l-\partial_l A_k$ are components of the local curvature form .

\section{Berry’s phase for coherent state in SU(2) group}

We calculate the Berry phase for a spin 1/2 particle in nonrelativistic quantum mechanics.  A coherent state for spin 1/2 particle is described by the following function [5]:

\begin{eqnarray}
\psi=
\left (
\begin{array}{c}
\cos\frac{\theta}{2}e^{-i\phi} \\ sin\frac{\theta}{2}
\end{array}
\right)
\end{eqnarray}

This eigenfunction are normalized on unit,

\begin{eqnarray}
\langle \phi | \phi \rangle=1.
\end{eqnarray}

The corresponding solution of the Schrodinger equation (1) is

\begin{eqnarray}
\psi=e^{i\Theta}\phi
\end{eqnarray}
where the phase $\Theta$  satisfies Eq. (7). Component of the local connection form $ A_k= i\langle \phi | \partial_k \phi \rangle$  for the eigenstate $\phi$   are easily calculated 

\begin{eqnarray}
A_{|\lambda|}=0,A_{\theta}=0, &  & {  } A_{\phi}=cos^2\frac{\theta}{2}
\end{eqnarray}

The respective local form of the curvature has only two nonzero components:

\begin{eqnarray}
F_{\theta\phi}=-F_{\phi\theta}=-\frac{1}{2}sin\theta
\end{eqnarray}

Now we calculate the Berry phase for a closed curve in the parameter space $ \lambda=\lambda(t) $ ,

\begin{eqnarray}
\Theta_B&=&\oint_{\lambda} d\lambda^k A_K=\frac{1}{2}\int\int_S d\lambda^k\times d\lambda^l F_{kl} \nonumber\\
&=&\int\int_S d\theta\times d\phi F_{\theta\phi}=-\frac{1}{2}\int\int d\theta \times d\phi sin\theta \nonumber\\
&=& -\frac{1}{2}\int (1-cos\theta)d\phi=-\frac{1}{2} \Omega(\lambda)
\end{eqnarray}

Where S is a surface in $R^3$  with the boundary $\lambda(t)$  and $\Omega(\lambda)$  is the solid angle of a surface S as it looks from the origin of the coordinate system.
This result does not depend on how parameters   depend on time [6].

We also calculate the Berry phase for a spin-1 particle in SU(2) in nonrelativistic quantum mechanics. Coherent state for spin-1 in real parameter is in the following form [5]:

\begin{eqnarray}
\psi=
\left (
\begin{array}{c}
e^{i\phi}sin^2\frac{\theta}{2} \\\frac{1}{\sqrt{2}}sin\theta \\ e^{-i\phi}cos^2\frac{\theta}{2}
\end{array}
\right)
\end{eqnarray}

If we consider solution of the Schrödinger equation (1) similar to equation (16), then component of the local connection form $ A_k=i\langle \phi | \partial_k \phi \rangle$   for the eigenstate  $\phi$  are easily calculated 

\begin{eqnarray}
A_{|\lambda|}=0,A_{\theta}=0, &  & {  } A_{\phi}=cos\theta
\end{eqnarray}
And components of the local form of the curvature are

\begin{eqnarray}
F_{\theta\phi}=-F_{\phi\theta}=-sin\theta
\end{eqnarray}

Now we calculate the Berry phase for a closed curve in the parameter space $ \lambda=\lambda(t) $ ,

\begin{eqnarray}
\Theta_B&=&\oint_{\lambda} d\lambda^k A_K=\frac{1}{2}\int\int_S d\lambda^k\times d\lambda^l F_{kl} \nonumber\\
&=&\int\int_S d\theta\times d\phi F_{\theta\phi}=-\int\int d\theta \times d\phi sin\theta \nonumber\\
&=& -\int (1-cos\theta)d\phi=-\Omega(\lambda)
\end{eqnarray}
where S is a surface in $R^3$  with the boundary $\lambda(t)$  and $\Omega(\lambda)$  is the solid angle of a surface S as it looks from the origin of the coordinate system.

 \section{Berry’s phase for coherent state in SU(3) group}

We calculate the Berry phase for a spin-1 particle in SU(3) in nonrelativistic quantum mechanics. Coherent state in real parameter in this group is in the following form [7]:

\begin{eqnarray}
\psi=
\left (
\begin{array}{c}
e^{i\phi}(e^{-i\gamma}sin^2\frac{\theta}{2}cosg-e^{i\gamma}cos^2\frac{\theta}{2}sing) \\
\frac{sin\theta}{\sqrt{2}}(e^{-i\gamma} cosg+e^{i\gamma}sing) \\ 
e^{-i\phi}(e^{-i\gamma}cos^2\frac{\theta}{2}cosg-e^{i\gamma}sin^2\frac{\theta}{2}sing)
\end{array}
\right)
\end{eqnarray}

This eigenfunction expression in such a way that they are normalized on unit,

\begin{eqnarray}
\langle \phi | \phi \rangle=1.
\end{eqnarray}

Similar to equation (16), then component of the local connection form  $A_k=i\langle \phi | \partial_k \phi \rangle$  for the eigenstate $\phi$    are easily calculated 

\begin{eqnarray}
A_{\theta}=0,A_g=0, &  & {  } A_{\phi}=cos\theta cos2g, A_{\gamma}=cos2g
\end{eqnarray}

Now we calculate the Berry phase for a closed curve in the parameter space ,

\begin{eqnarray}
\Theta_B&=&\oint_{\lambda} d\lambda^k A_K=\frac{1}{2}\int\int_S d\lambda^k\times d\lambda^l F_{kl} \nonumber\\
&=&\int\int d\theta d\phi F_{\theta\phi}+\int\int d\theta dg F_{\theta g}+\int\int d\theta d\gamma F_{\theta \gamma} \nonumber\\
& &+\int\int dgd\phi F_{g\phi}+\int\int d\gamma d\phi F_{\gamma\phi}+\int\int dgd\gamma F_{g\gamma} \nonumber\\
&=& \int\int d\theta d\phi cos2gsin\theta-2\int\int dgd\phi sin2gcos\theta-2\int\int dgd\gamma sin2g \nonumber\\
&=&\int (cos2g-cos\theta)d\phi-\int (1-cos2g)d\gamma
\end{eqnarray}

In the above relation if we set $g=0$ , obtain Berry phase in SU(2) group.

\section{Discussion}

Geometric phases are important in quantum physics and are now central to fault tolerant quantum computation. We have presented a detailed analysis of geometrical phase that can arise within general representations of coherent states in real parameterization in SU(2) and SU(3) groups. As coherent state in SU(3) group with $g=0$  convert to coherent state in SU(2), Berry phase also change in similar method. We can continues this method to obtain Berry phase in SU(N) group, where $N\geq4$  . we can also obtain Berry phase from complex variable base ket, we conclusion that result in two different base ket is similar. Berry phase application in optic, magnetic resonance, molecular and atomic physics [8,9] .

\end{document}